\documentstyle[sprocl,epsfig]{article}

\def\Journal#1#2#3#4{{#1} {\bf #2}, #3 (#4)}

\def\NPB{{\em Nucl. Phys.} B}
\def\PLB{{\em Phys. Lett.}  B}
\def\PRL{\em Phys. Rev. Lett.}

\def\be{\begin{equation}}
\def\ee{\end{equation}}
\def\beq{\begin{equation}}
\def\eeq{\end{equation}}
\def\bea{\begin{eqnarray}}
\def\eea{\end{eqnarray}}

 \def\PgSm{\ifmmode{\rm \Sigma^-}
           \else${\rm \Sigma^-}$\fi}
 \def\PgXm{\ifmmode{\rm \Xi^-}
           \else${\rm \Xi^-}$\fi}
 \def\Pe{\ifmmode{\rm e}
         \else${\rm e}$\fi}
 \def\Pgpm{\ifmmode{\rm \pi^-}
           \else${\rm \pi^-}$\fi}
 \def\gevc1{\ifmmode\mathrm{ GeV/{\mit c}}
          \else$\mathrm{ GeV/{\mit c}}$\fi}

\begin{document}

\title{Hyperon Radii}

\author{B. POVH}

\address{Max-Planck-Institut f\"ur Kernphysik,\\ 
Post Box 103980\\
D-69029 HEIDELBERG\\E-mail: b.povh@mpi-hd.mpg.de} 




\maketitle\abstracts{ A survey of the experimental data 
on the charge and strong hadronic radii is given. A 
new, however preliminary value of the $\Sigma^-$ charge radius
allows to include also hyperons in the systematics.
The presented data on the hadronic radii give the first hint
of their dependence on the strangeness content. An estimate
of the constituent quark size is given.}

\section{Introduction}

Constituent quark models, nonrelativistic and semi-relativistic,
reproduce well the static properties and low excitations of hadrons.
Probably the most convincing support of the constituent quark concept
is the impressive agreement between the model predictions for the
magnetic moments and the data. The data on the baryon octet, six
measured magnetic moments in total, give a sufficient constraint to
the predictions of the model to make the concept of the constituent
quark credible. It has essentially two free parameters, the light and
the strange quark mass.  Here we do not count the parameters of the
quark-quark potential which is assumed to be such that the kinetic and
the potential energy cancel each other.

Hadronic radii can supply us with further information on the
constituent quark properties. The sizes of the hadrons are given 
by the sizes of the constituent quarks and the confinement
forces. Again, if the charge radii of the hyperons were determined,
we would get five independent data points on the dependence of the hyperon
radii on the content of the strangeness: four by replacing the
light quarks in the nucleon by the strange ones, the fifth by 
comparing the $\Sigma^+$ and $\Sigma^-$ which should have 
different charge radii because the u- and d-quarks in the two
hyperons lead to a different charge distribution.
 
The aim of this paper is twofold. Firstly, by presenting the determination
of the charge radius of $\Sigma^-$ we show that a systematic
measurements of hyperon radii is feasible with present
experimental techniques and existing hyperon beams.
Secondly, by comparing charge and strong radii, 
the latter obtained from the hadronic cross sections, we demonstrate the
equivalence between the two. 

The presented data on the hadronic radii
give the first hint of their dependence
on the strangeness content and on the quark sizes.  

\section{$\Sigma^-$-Electron Scattering}

The first measurement of charge radii of unstable particles,
pion and kaon,
was performed at CERN in the early 80's by 
Amendolia {\em et al.}\cite{ampi86,amkm86}.
Negative pions and kaons of 300 GeV were scattered on electrons of a
liquid hydrogen target. The angles of the scattered mesons and
recoil electrons were measured by a set of wire chambers, the 
momenta by the following magnetic spectrometer. 

\begin{figure}[htbp]
  \begin{center}
    \psfig{file=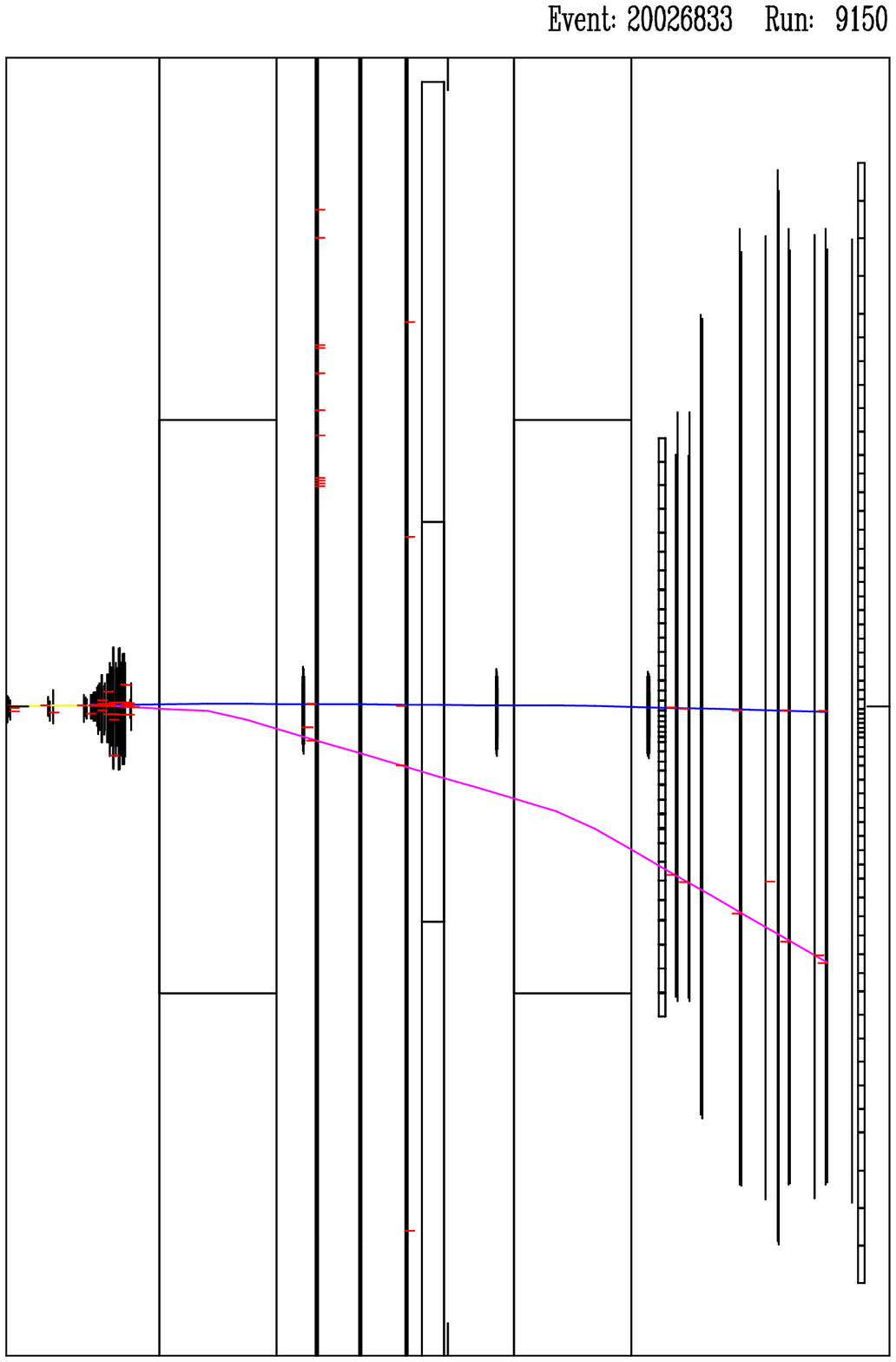,width=7cm,height=6cm}
    \hspace{-1.2cm}\psfig{file=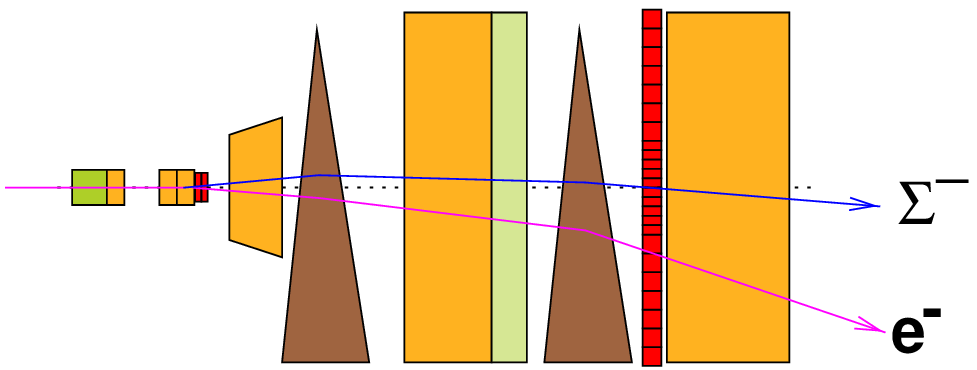,width=8cm,height=2cm}
    \caption{Event display of an elastic \PgSm-\Pe\ scattering event in the SELEX detector.}
    \label{fig:event}
  \end{center}
\end{figure}

Intense hyperon beams operate at the highest possible energies in
order to increase the decay length of the hyperons. A high hyperon
energy is also vital for the measurement of the electromagnetic form
factors in order to obtain scattering events with a sufficiently high
$Q^2$ bite to deduce at least the charge radius, the first moment of
the charge distribution. The beam used by the SELEX collaboration at
Fermilab was tuned to 600 GeV. At this momentum of $\Sigma^-$ the
energy of the recoil electrons range from 0 to 180 GeV, while the
largest angle for the scattered $\Sigma^-$ is 0.5 mrad and for the
recoiling electrons 8 mrad in the momentum acceptance of the
spectrometer.  A typical scattering event is shown in Fig.
\ref{fig:event}.


In Fig. \ref{fig:event}, only the part of the SELEX experiment relevant for the 
identification of the scattering events is shown. The momenta and 
angles of the incoming $\Sigma^-$ were determined by reconstructing
the beam tracks in the set of silicon strip detectors in front of the
target, the scattering angles by the detector behind the target. The 
momenta of the scattered particles were obtained by the reconstruction 
of the particle trajectories in the following two magnetic spectrometers.
Silicon strip detectors, 5 cm in diameter, were positioned
behind the first and in front of and behind the second magnet. 

An average uncertainty in the determination of the scattering angle
was 30~$\mu\mbox{rad}$, and the momentum for the $\Sigma^-$ was determined
by an accuracy of $\approx 1\%$. The four-momentum transfer $Q^2$
was calculated by using the scattering angles of the $\Sigma^-$,
of the recoiling electron and the $\Sigma^-$ momentum.
The electron momentum was not fully used in the analysis because
of large radiation corrections to be applied to it.

\begin{figure}[htbp]
  \begin{center}
    \psfig{file=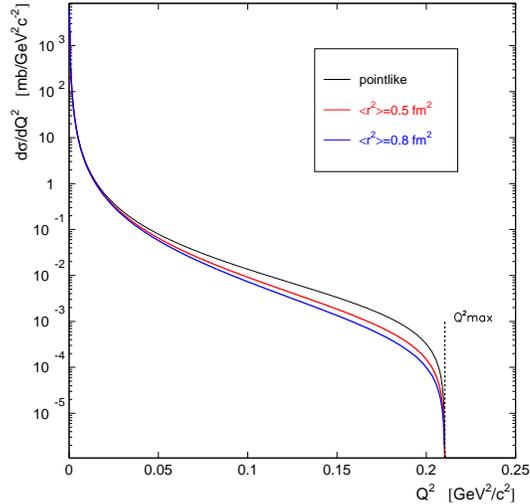, width=7cm}
    \caption{Differential cross section of elastic \PgSm-\Pe\ scattering at 650 \gevc1.}
    \label{fig:xsec}
  \end{center}
\end{figure}

The experimentally obtained cross sections have been fitted to 
the following formula:
\beq
\Biggl(\frac{d\sigma}{dQ^2}\Biggr)_{exp}=\Biggl(\frac{4\pi\alpha}
{Q^4}\Biggl(1-\frac{Q^2}{Q^2_{max}}\Biggr)\Biggr)_{Mott}\cdot F^2(Q^2).
\eeq
The form factor $F(Q^2)$ depends on the electric and the magnetic
form factor. As generally used in the electron-proton scattering and
also for the $\Sigma^-$ the Sachs form factors with 
the empirical dipole dependence on $Q^2$ are assumed:

\beq
G_E(Q^2)=\frac{1}{\kappa -1} G_M(Q^2)=\Bigl(1+\frac{1}{6}Q^2\langle r_{ch}^2\rangle\Bigr)^{-2}.
\eeq
In the case of $\Sigma^-$ the anomalous magnetic moment $\kappa=-\ 0.16$
is very small.
In Fig. \ref{fig:xsec}, the $Q^2$ dependence of the cross section is shown for
different radii.

Because of the large variation in data for the proton charge radius
(see Tab.~1) from various experiments, indicating uncontrolled
systematic errors, it was felt that the $\Sigma^-$ charge radius
should be compared to the proton radius obtained in the same experiment.
The preliminary value for the $\Sigma^-$ radius is $\langle
r^2_{ch}\rangle_{\Sigma^-}= (0.60 +
0.08(stat.)+0.08(syst.))\mbox{fm}^2$ and for the proton $\langle
r^2_{ch}\rangle_p= (0.84+0.09(stat.)+0.06(syst.))\mbox{fm}^2$.  In
spite of the large errors the results strongly indicate that the
$\Sigma^-$ radius is smaller than that of the proton.

\section{Charge vs. Strong Radii}

\begin{figure}[htbp]
  \begin{center}
    \psfig{file=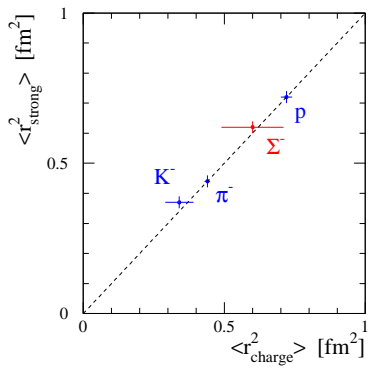,width=5cm}
    \caption{Strong versus measured electromagnetic charge radii of hadron.}
    \label{fig:comp}
  \end{center}
\end{figure}

A meson interacts with a nucleon like a colour dipole. A baryon
can also be viewed as a coherent sum of colour dipoles.
Therefore, it is expected that the total cross sections of hadrons
on a nucleon are proportional to the $\langle r^2\rangle_h$ of the hadron.
It was shown by Povh and H\"ufner \cite{PovhHuef1} that
at sufficiently high energies, $\sqrt s \ge 20$GeV, the total 
cross section of hadrons on the nucleon 
are given by

\beq
\frac{\sigma(h,p)}{\sigma(p,p)}\simeq \frac{\langle r^2\rangle_h}{\langle r^2\rangle_p},
\eeq
providing the cross sections are measured at the same energies.
The total hadronic cross sections grow logarithmically with energy.
This is the consequence of the gluon halo of which softer and 
softer components interact with increasing energy (H\"ufner and Povh \cite{PovhHuef2}).
It is not possible to define an energy independent strong radius, but a good
fit to the data is obtained assuming 
\beq
\langle r^2(s)\rangle_{strong}=\langle r^2\rangle_{ch}\cdot\Bigl(\frac{s}{s_0}\Bigr)^{\Delta}.
\eeq
$\Delta$ has a value of about $0.1$. The correspondence between the
strong and charge radii is shown in Fig. \ref{fig:comp}. The total cross sections
and the strong radii deduced from them are better and more abundantly
measured than the charge radii. Therefore, in the following discussion 
we will combine the information from the two sources in order to 
discuss the flavour dependence of the hadronic sizes.

\section{Flavour Dependence of the Hadronic Radii}

The charge radius gives the first moment of the charge distribution,
the strong, as defined above, the first moment of the mass distribution.
The two are in general not the same because the quarks have different
electric charges but the same strong coupling. Nevertheless, for all
the hadrons quoted in Tab.~1, the two radii are the same if
properly scaled. 
This scaling is done by adjusting the strong radius of proton to be
equal to its charge radius.
In the following discussion we will just talk 
about the hadronic radii. In fact, the only exception in the family
of hyperons is $\Sigma^+$ which is expected to have the charge
radius larger than the strong one. This is because the two {\sl up} quarks
with the summed charge of 4/3 units surround the {\sl strange} quark
that very likely has smaller extention than the light quarks.

\begin{table}[htbp]
  \caption{Comparison of measured electromagnetic charge radii to strong interaction radii}
  \begin{center}
    \begin{tabular}{|l|c|c|}
      \hline
      \vrule height10pt width0pt
      &\multicolumn{2}{c|}{mean square radius [fm$^2$] } \\
      \cline{2-3}
      \vrule height10pt width0pt
       & electromagnetic interaction & strong interaction \\      \hline
      \vrule height10pt width0pt
       p      & $0.74\pm 0.02$   \cite{bork80}  &            \\
       p      & $0.79\pm 0.03$   \cite{udem97}  &            \\
       p      & $0.72\pm 0.01$   \cite{merg96}  & $0.72\pm 0.02$           \\
       n      & $-0.11\pm 0.03$  \cite{koes76}  &             \\ 
       \PgSm  & $0.9\pm 0.5$     \cite{wa89}    & $0.62\pm 0.02$ \\
      
       \PgXm  &                  & $ 0.54\pm 0.02$ \\
       \Pgpm  & $0.44\pm 0.01$   \cite{ampi86}  & $0.43\pm 0.02$  \\
       $K^-$  & $ 0.34\pm 0.05$  \cite{amkm86}  & $0.37\pm 0.02$  \\
       $K^0$  & $-0.054\pm0.026$ \cite{molz78}  &            \\
      \hline
    \end{tabular}
  \end{center}
\end{table}
Let us denote the mass distribution of the light quarks with
$\Phi (q)$ and for the strange quark with $\Phi (s)$. The
radii of the hadrons in the right column of the Table 1.
can be written as:
\begin{eqnarray}
&\langle r^2\rangle_p&=\int \Phi_h (q)r^2 d^3r\\
&\langle r^2\rangle_{\Sigma^-}&=\int [2/3 \Phi_h(q)+1/3 \Phi_h(s)]r^2 d^3r\\
&\langle r^2\rangle_{\Xi^-}&=\int [1/3 \Phi_h (q) +2/3 \Phi_h (s)] r^2 d^3 r\\
&\langle r^2\rangle_{\pi^-}&=\int 2/3 \Phi_g(q) r^2 d^3\\
&\langle r^2\rangle_{K^-}&=\int 1/3[\Phi_g(q) +\Phi_g(s)] r^2 d^3r.
\end{eqnarray}
Here we assumed that in the first approximation $\Phi_h(q)$ and
$\Phi_h(s)$ do not depend strongly on the number of the strange
quarks in the hyperon. On the other hand, $\Phi_g(q)$ and $\Phi_g(s)$,
the density distributions of the Goldstone particles
$\pi^-$ and $K^-$ certainly differ from those of the hyperons in the 
contribution coming from the relative motion of the quarks. 

The differences of the square radii:
\begin{eqnarray}
  \langle r^2\rangle_p-\langle r^2\rangle_{\Sigma^-} \\
 \langle r^2\rangle_{\Sigma^-}-\langle r^2\rangle_{\Xi^-} \\
 \langle r^2\rangle_{\pi^-}-\langle r^2\rangle_{K^-} 
\end{eqnarray}
have,
within the experimental errors, the same value, 
$\Delta \langle r^2\rangle=0.1\pm0.03$.
Thus we can write
\beq
\int [\Phi_h(q)-\Phi_h(s)]r^2d^3r \simeq \int [\Phi_g(g)-\Phi_g(s)] r^2d^3r=
0.3\pm0.09 fm^2.
\eeq
This result implies a very strong radius dependence on the quark
mass. If the density distribution were only the consequence of the 
relative motion of the quarks in the hyperon the mass dependence 
of the radius were rather weak. The fact that 
the $\Delta\langle r^2\rangle$ are approximately equal in the hyperons and
the Goldstone bosons suggests that the major part of this difference
is due to the different intrinsic radii of the constituent light and strange
quarks. 

It was pointed out by Povh and H\"ufner \cite{PovhHuef3} that in the
nonrelativistic quark model the hadronic excitations and the hadronic
radii are consistently described if one attributes an apparent radius
to each constituent quark,\\
$\langle r^2\rangle_{u,d}=0.36~$fm$^2$ for the light
and $\langle r^2\rangle_s=0.16~$fm$^2$ for the strange quark.\\
This result is not surprising. The radius of a nonrelativistic quark
is just the Compton wave length of the quark, $\langle r^2\rangle^{1/2} \approx
\frac {1}{m_q}$, which is rather obvious for a confined particle.
The virtual creation of quark-antiquark
pairs in the confining field leads to a fluctuation ("Zitterbewegung")
in the position coordinate of the quark, which appears
in the nonrelativistic Schr\"odinger picture as an apparent size
of the moving quark, with a value close to the Compton wave length.
This effect has been calculated by Hayne and Isgur \cite{HayIsg}
for the non-relativistic quark model. 

In conclusion we want to point out that the same quark masses used to
optimise the magnetic moments and the low lying excited states of the
hyperons do give an excellent result for the hadronic radii if one assumes an
intrinsic radius of the constituent quark to have the value of its Compton
wave length.


\section*{Acknowledgments}
I would like to thank my colleagues H. Kr\"uger, J. Simon and
G. Dirkes for supplying me with the results of the SELEX experiment.  


\end{document}